\documentclass[12pt]{JHEP3}

\usepackage{epsfig}

\def\spose#1{\hbox to 0pt{#1\hss}}
\def\ltapprox{\mathrel{\spose{\lower 3pt\hbox{$\mathchar"218$}}
 \raise 2.0pt\hbox{$\mathchar"13C$}}}
\def\gtapprox{\mathrel{\spose{\lower 3pt\hbox{$\mathchar"218$}}
 \raise 2.0pt\hbox{$\mathchar"13E$}}}

\title{Non-commutative Principal Chiral Models}

\author{Stefano Profumo \\
        Scuola Internazionale Superiore di Studi Avanzati \\
	Via Beirut 2-4, I-34014 Trieste, Italy \\ 
	E-mail: \email{profumo@sissa.it} 
}

\abstract{Twisted Eguchi-Kawai reduced chiral models are shown to
be formally equivalent to a $U(1)$ non-commutative parent theory.
The non-commutative theory describes the vacuum dynamics of the
non-commutative charged tachyonic field of a brane system. To make
contact with the continuum non-commutative theory, a double scaling
large $N$ limit for the reduced model is required. We show a
possible limiting procedure, which we propose to investigate
numerically. Our numerical results show substantial consistency with the outlined procedure. 
}


\preprint{}

\begin{document}

\section{Introduction}
Twisted Eguchi-Kawai (TEK) reduced models \cite{reduced} provide a
non-perturbative definition of certain non-commutative field
theories \cite{AIKKT,M,Metal}. It has been shown that the $U(N)$
gauge theory with gauge fields obeying twisted boundary conditions
over the non-commutative torus $\mathbb{T}^D_{\Theta}$ is
equivalent to a $U(\tilde{N})$ gauge theory, with $\tilde{N}$
suitably chosen, over the non-commutative torus
$\mathbb{T}^D_{\Theta^\prime}$ with gauge fields obeying periodic
boundary conditions \cite{Metal}. This is a consequence of a more
general fact, known as \emph{Morita equivalence}.

We would like to propose an application of the formalism of ref.
\cite{Metal,szabo} to principal chiral models, possibly providing
numerical evidence, in order to show that two-dimensional TEK
reduced chiral models can be considered as a non-perturbative
definition of a non-commutative field theory.

Explicitly, we are going to show the equivalence of TEK principal
chiral model with symmetry group $U(N)$ to a non-commutative $U(1)$
lattice theory compactified on a torus with periodic boundary
conditions. We will describe the corresponding non-commutative
theory, i.e. its action and symmetries, coupling constant and
dimensionful non-commutativity parameter. We will eventually try to
define a procedure which can lead to a sensible continuum limit,
and therefore define the set of values of the coupling constants
and of the symmetry parameter $N$ of the original TEK theory
relevant for a numerical check of the consistency of the whole approach.

The reduced model, in this context, should be considered in a
different limit from the original planar one. In the planar limit
one has to send $N\rightarrow\infty$ while keeping the lattice
coupling constant $\beta$ fixed. The continuum limit is then
reached as one sends $\beta\rightarrow\infty$, because the RG relation
$a\propto e^{-c\cdot\beta}$ exists between the lattice
spacing $a$ and the lattice coupling constant $\beta$. The limit
is called planar since the large $N$ dominant Feynman graphs are, in
this limit,
 planar, i.e. they can be drawn on a Riemann surface of genus $g=0$.

For the TEK model to reproduce the corresponding non-commutative
field theory we must consider a different limiting procedure to
approach the continuum limit. This procedure is called \emph{double
scaling limit}, since $\beta$ and $N$ must be sent to infinity in
a correlated manner. We know that in the non reduced original
field theory this corresponds to taking into consideration also
non-planar graphs, analogously to what happens in matrix models of
$2D$ gravity: contributions from higher genus topologies imply
a higher symmetry in the problem, which enables to make
contact with the theory one wants to reproduce. Quite the same
thing happens in non-commutative field theory, since the
interaction vertex is invariant only up to cyclic permutations of
the momenta, and therefore one needs to keep track of the cyclic
order in which lines emanate from vertices in a given Feynman
diagram. Non-commutative Feynman graphs are thus ribbon graphs
drawn on Riemann surfaces of  particular genus, in complete
analogy with what happens in the ordinary large $N$ limit of field
theories. Non-planar contributions thus naturally arise in the
present context of non-commutative field theories.

Principal chiral models in $D=2$ are since long \cite{GS1981}
known to be in a sense\footnote{This correspondence is exact in $D=1$ and
$D=2$ respectively.} a simpler counterpart of lattice Yang-Mills
theory in $D=4$. Also as far as the reduced TEK models of the
two theories are concerned, the case of chiral models is more
tractable, and even numerical simulations are easier and more
conclusive. For this reason we consider principal chiral model as
an important test for the non-commutative interpretation of TEK
reduced models.

Furthermore, the non-commutative theory arising from the TEK
reduced principal chiral model turns out to possess exactly the
action needed to describe the vacuum dynamics of the
non-commutative charged tachion of a certain brane system
\cite{Sochi}.

We begin with a general description of non-commutative field
theories, in order to fix the notation, and describe the
particular non-commutative field theory we are interested in, i.e.
$2D$ principal chiral models. We then pass to a short analysis of
the reduced models and of the consequences of our numerical
results. In the final section we demonstrate the equivalence
between TEK reduced chiral models and non-commutative $U(1)$
principal chiral models on the lattice, indicating the procedure
needed to approach a sensible continuum limit. This will
eventually lead to a concrete proposal of a numerical study of the
theory reread in this new light. The physical interpretation of
the presented non-commutative theory is then briefly sketched. We include the results of 
our Monte Carlo simulations,
which directly enforce the correctness of the whole procedure.


\section{Non-commutative Field Theories}\label{sezformalismo}

We will briefly recall the so called Weyl quantization procedure
for field theories on non-commutative spaces. Let us consider for
simplicity a scalar field theory on Euclidean $\mathbb{R}^D$,
defined by some action $S[\phi]$, and whose partition function is
as usual

\begin{equation}
Z \ = \ \int \mathcal{D}\phi \ e^{-S[\phi]}.
\end{equation}
In order to pass from ordinary to non-commutative space-time, we
replace the local coordinates $x_\mu$ by hermitian operators
$\hat{x}_\mu$ which have the following commutations relations:

\begin{equation} [\hat{x}_\mu,\hat{x}_\nu] \ = \ i\theta_{\mu\nu}, \end{equation}
where $\theta_{\mu\nu}=-\theta_{\nu\mu}$ is a real valued
anti-symmetric matrix with the dimensions of length squared.
Implementing the substitution $x_\mu\rightarrow\hat{x}_\mu$ we
obtain the \emph{Weyl operator} $\hat{\phi}$ corresponding to the field
$\phi$.\\
One can define an operator which transforms a field into its Weyl
operator:

\begin{equation}\label{delta}
\begin{array}{c}
\hat{\phi}=\int \mathrm{d}^Dx\phi(x)\hat{\Delta}(x),\\
\\
\hat{\Delta}(x) \ = \ \int
\frac{\mathrm{d}^Dk}{(2\pi)^D}e^{ik_\mu\hat{x}^\mu}e^{-ik_\mu
x^\mu}.
\end{array}
\end{equation}
It is possible to introduce an anti-hermitian derivation
$\hat{\partial}_\mu$ through the commutations relations

\begin{equation}
\big[\hat{\partial}_\mu,\hat{x}_\nu\big]=\delta_{\mu\nu},\qquad
\big[\hat{\partial}_\mu,\hat{\partial}_\nu\big]=0.
\end{equation}
The operator $\hat{\partial}_\mu$ obviously satisfies the following
relations:

\begin{equation}
\begin{array}{c}
\big[\hat{\partial}_\mu,\hat{\phi}\big] \ = \ \int
\mathrm{d}^Dx \ \hat{\partial}_\mu\phi(x)\hat{\Delta}(x),\\
\\
\big[\hat{\partial}_\mu,\hat{\Delta}(x)\big] \ = \
\hat{\partial}_\mu\hat{\Delta}(x).
\end{array}
\end{equation}
This implies that, given the generator of the translations
$e^{v_\mu\hat{\partial}^\mu}$, $v_\mu\in\mathbb{R}$, which
satisfies

\begin{equation}
e^{v_\mu\hat{\partial}^\mu}\hat{\Delta}(x)e^{-v_\mu\hat{\partial}^mu}=\hat{\Delta}(x+v),
\end{equation}
the operation $\mathrm{Tr}\hat{\Delta}(x)$ is independent of $x$
for any trace on the algebra of operators. Therefore the
integration of the fields on the space-time is represented by

\begin{equation} \mathrm{Tr}\hat{\phi} \ = \ \int \mathrm{d}^Dx \ \phi(x),
\qquad \mathrm{Tr}\hat{\Delta}(x)=1. \end{equation}
This enables to define the inverse of the correspondence between
fields and operators:

\begin{equation}\label{inversa} \phi(x) \ = \
\mathrm{Tr}\left(\hat{\phi}\hat{\Delta}(x)\right). \end{equation}
In conclusion, in order to pass from ordinary to  non-commutative
field theory, we need to implement the following substitutions:

\begin{equation} \phi\rightarrow\hat{\phi},\qquad
\partial_\mu\phi\rightarrow\big[\hat{\partial}_\mu,\hat{\phi}\big].
\end{equation}
The substantial difference between the two theories comes from the
definition of products of fields. In the non-commutative case one
has, for $\hat{\phi}_3=\hat{\phi}_1\hat{\phi}_2$

\begin{equation}\label{contstarprod}
\begin{array}{c}
\phi_3(x)=\mathrm{Tr}\left(\hat{\phi}_1\hat{\phi}_2\hat{\Delta}(x)\right)=\\
\\
=\frac{1}{\pi^D|\mathrm{det}\theta|}\int\int\mathrm{d}^D \
y\mathrm{d}^Dz\phi_1(y)\phi_2(z)e^{-2i(\theta^{-1})_{\mu\nu}
(x-y)_\mu(x-z)_\nu}=\\
\\
=\phi_1(x)\exp\left(\frac{i}{2}\overleftarrow{\partial_\mu}\theta_{\mu\nu}
\overrightarrow{\partial_\nu}\right)\phi_2(x)
\equiv \phi_1(x)\star\phi_2(x)=\\
\\
=\phi_1(x)\phi_2(x)+\sum_{n=1}^\infty\left(\frac{i}{2}\right)^n\frac{1}{n!}\theta^{i_1j_1}\cdots\theta^{i_nj_n}
\partial_{i_1}\cdots\partial_{i_n}\phi_1(x)\partial_{j_1}\cdots\partial_{j_n}\phi_2(x),
\end{array}
\end{equation}
which defines the \emph{star} or \emph{Moyal product} of the fields.

\subsection{Non-commutative $U(N)$ Principal Chiral Models}

The theory we would like to study is defined in a $D$ dimensional
Euclidean space by the following action, multiplied by a suitable
coupling constant

\begin{equation} S \ = \ \beta N\int \ \mathrm{d}^Dx \ \mathrm{tr}_{(N)}\left(\partial_\mu
U(x)\partial^\mu U^\dag(x)\right), \end{equation}
where the sum over $\mu=1,2,...,D$, and a flat Euclidean metric are intended\footnote{In what
follows we will restrict to the case $D=2$.}. The unitary matrices
$U_{ij}(x)$, with $i,j=1,2,...,N$ satisfy the condition

\begin{equation} \left(U^\dag(x)U(x)\right)_{ij}=\delta_{ij} \qquad \forall \
x. \end{equation}
The partition function is defined over the usual Haar measure

\begin{equation} Z \ = \ \int \ \mathrm{d}U \ e^{-S[U]}. \end{equation}
The action is naturally invariant under the transformations (where
\mbox{$V_L,V_R \in U(N)$})

\begin{equation} U\rightarrow V_LU, \qquad U\rightarrow UV_R. \end{equation}
In the present formalism, one needs to define the Weyl operators
as

\begin{equation} \hat{U} \ \equiv \ \int \mathrm{d}^Dx  \ \hat{\Delta}(x)\cdot
U(x), \end{equation}
where the operator $\hat{\Delta}(x)$ is the one defined in
(\ref{delta}). The action is rewritten as

\begin{equation}\label{weylPCM}
S \ = \ \beta N \ \mathrm{Tr} \ \Big\{ \ \mathrm{tr}_{(N)}\left(
\Big[\hat{\partial_\mu},\hat{U}\Big]\Big[\hat{\partial^\mu},\hat{U^\dag}\Big]\right) \ \Big\},
\end{equation}
where $\mathrm{Tr}$ is the trace operator over coordinates, while
$\mathrm{tr_{(N)}}$ is the (finite-dimensional) trace in the
fundamental representation of the $U(N)$ group.\\
The corresponding non-commutative fields $\mathcal{U}(x)$, defined
through the inverse transformation (\ref{inversa}), satisfy the
star-unitarity condition\footnote{since $\det (f\star g)\neq \det
(f) \star \det (g)$ we cannot take $SU(N)$ as the symmetry group
of the theory.}

\begin{equation}\label{starunit}
\mathcal{U}(x) \ \star \ \mathcal{U}(x)^\dag \ = \
\mathcal{U}(x)^\dag \ \star \ \mathcal{U}(x) \ = \ \mathbb{I}_N.
\end{equation}
The action is given by the obvious translation of eq.
(\ref{weylPCM})

\begin{equation}\label{fieldaction}
S \ = \ \beta N \int\mathrm{d}^Dx \ \mathrm{tr}_{(N)}\left(
\partial_\mu\mathcal{U}(x)\star\partial^\mu\mathcal{U}(x)^\dag\right).
\end{equation}
The invariance of the action naturally reads, owing to cyclicity
of the trace $\mathrm{tr}_{(N)}$,

\begin{equation} \mathcal{U}(x)\rightarrow g_L \ \mathcal{U}(x) \qquad \mathcal{U}(x)\rightarrow \mathcal{U}(x)
\ g_R, \end{equation}
where the $N\times N$ matrices $g_L,g_R$ are ordinary unitary
matrices, therefore satisfying $gg^\dag=\mathbb{I}_N$.

\section{The reduced model}

The principal chiral model, in $D=2$, is defined on the lattice
via the usual substitution of the derivative with a finite
difference

\begin{equation}
\partial_\mu U(x) \ \rightarrow \
\frac{U_{x+a\hat{\mu}}-U_{x}}{a},
\end{equation}
and the resulting action reads

\begin{equation}\label{azionelattice}
S=-\beta N\sum_{x}\sum_{\mu=1,2}\mathrm{tr}_{(N)} \
\Big[U_xU^\dag_{x+\mu}+U_{x+\mu}U^\dag_x\Big].
\end{equation}
The naive Eguchi-Kawai reduction prescription $U_x\rightarrow
U$ is clearly not applicable in this context. Instead, one can
resort to the TEK prescription, which is defined as

\begin{equation}\label{gammini}
U_{(x_1,x_2)} \ \rightarrow \
\Gamma_1^{x_1}\Gamma_2^{x_2}U(\Gamma_2^\dag)^{x_2}(\Gamma_1^\dag)^{x_1},
\end{equation}
where the \emph{twist matrices} $\Gamma_\mu$ obey the Weyl-'t
Hooft algebra

\begin{equation}\label{WtH}
\Gamma_\mu\Gamma_\nu \ = \ \exp\Bigg[\frac{2\pi
i}{N}n_{\mu\nu}\Bigg]\Gamma_\nu\Gamma_\mu,
\end{equation}
where $N$ is the parameter of the symmetry group $U(N)$ or $SU(N)$
(and thus of the matrix $U$) and $n_{\mu\nu}$ is an integer valued
antisymmetric tensor, whose generic form in $D=2$ is of course

\begin{equation}
n_{\mu\nu} \ \equiv \ \left(
\begin{array}{c c}
0 & M \\
-M & 0
\end{array}
\right) \ , \qquad M\in\mathbb{Z}.
\end{equation}
For a given $N$ and $M$ the solution to (\ref{WtH}) is provided,
up to global $SU(N)$ transformations, by the $N\times N$ shift and
clock matrices\footnote{The two matrices are a natural extension
of the known 't Hooft twist matrices \cite{thoofttwists}.}

\begin{equation}
\begin{array}{c}
S^{(M)}_{i,j} \ \equiv \ \delta_{i+M,j},\\
\\
C_{i,j} \ \equiv \ e^{\frac{2\pi i}{N}(i-1)}\delta_{i,j}.
\end{array}
\end{equation}
The two matrices $\Gamma_\mu$ will be given respectively by $S$
and $C$.\\
Applying the reduction prescription to the action
(\ref{azionelattice}) gives

\begin{equation}\label{azioneTEK}
S_{TEK}=-\beta N\sum_{\mu=1,2}\mathrm{Tr}\Big[U\Gamma_\mu
U^\dag\Gamma^\dag_\mu+h.c.\Big].
\end{equation}
We notice that the model shows two symmetries, namely:
\begin{enumerate}
\item $U \ \rightarrow \
\Gamma_1^{x_1}\Gamma_2^{x_2}U(\Gamma_2^{x_2})^\dag(\Gamma_1^{x_1})^\dag$;
\item $U \ \rightarrow \ z\cdot U, \qquad z\in\mathbb{Z}_N$.
\end{enumerate}
The first symmetry is reminiscent of the space-time translational
symmetry of the original model (indeed from (\ref{gammini}) it is
clear that the role of the $\Gamma_\mu$ is that of generators of
translations in the dual lattice of the reduced theory) while the
second represents the residual global symmetry $SU(N)\times SU(N)$
of the parent theory reduced to the center $\mathbb{Z}_N$ of the
algebra of $SU(N)$. In the case of a symmetry $U(N)\times U(N)$
the second
symmetry is of course a $U(1)$ symmetry.

The introduction of the TEK reduced model was originally motivated
by its supposed equivalence with the parent theory (defined by the
action of eq. (\ref{azionelattice})) in the large $N$ limit. This equivalence should
follow basically from two facts:
\begin{itemize}
\item[-] The Schwinger-Dyson (SD) equations of the reduced theory and
that of the parent theory are exactly the same in the large $N$ limit, up to terms which
are not invariant under the $\mathbb{Z}_N$ symmetry
\item[-] If in all regimes the symmetry is not spontaneously
broken, the two models possess exactly the same SD set of
equations, and given the same initial conditions should coincide
\end{itemize}
It has been found \cite{future} that the invoked equivalence holds
for the strong coupling region, while in the weak coupling the
reduced theory is manifestly different from its parent version.
Thus, although the second symmetry does not seem to be
spontaneously broken, the SD based argument for the equivalence is
not sufficient, at least in a certain regime, to completely
identify
the theory.

It therefore makes sense to ask whether the reduced principal
chiral model can provide a non perturbative definition of
\emph{another} theory. Our claim is that this theory can be
interpreted as the \emph{non-commutative $U(1)$ version of}
(\ref{azionelattice}), and that, in a suitable limit to be
defined, it reproduces its continuum limit, i.e. the one specified
in the Weyl operator notation by eq. (\ref{weylPCM}) and in the
non-commutative field notation by eq. (\ref{fieldaction}).

\section{Reduced chiral models and non-commutative chiral fields}

We will show in what follows how to map the TEK model into
the non-commutative version of the theory defined by the action
(\ref{azionelattice}). Given an integer valued vector
$\mathbf{k}=(k_1,k_2)$, we introduce the $N \times N$ matrices

\begin{equation}
J_k=\Gamma_1^{k_1}\Gamma_2^{k_2} e^{\pi i (n_{12})
k_2k_1/N}=\Gamma_1^{k_1}\Gamma_2^{k_2} e^{\pi i M k_2k_1/N}.
\end{equation}
The phase factor is  given to symmetrically order the product of
twist eaters. Incidentally, the $J_k$'s have the same algebraic
properties as the plane Weyl basis $e^{ik_i\hat{x}^i}$ for the
continuum non-commutative field theory on the torus \cite{szabo}.\\
The relevant properties of these matrices are that there are only
$N^2$ such matrices, owing to the periodicity properties

\begin{equation}\label{periodicity}
J_{N-k} \ = \ J_{-k} \ = \ J^\dag_k,
\end{equation}
and that they obey the orthonormality and completeness relations

\begin{equation}
\begin{array}{r c l}
\frac{1}{N}\mathrm{tr}_{(N)}\left(J_kJ_q^\dag\right) & = &
\delta_{k,q(\mathrm{mod} \ N)},\\
\\
\frac{1}{N}\sum_{k\in\mathbb{Z}_N^2}(J_k)_{\mu\nu}(J_k)_{\lambda\rho}
& = & \delta_{\mu\rho}\delta_{\nu\lambda}.
\end{array}
\end{equation}
They therefore form a basis for the linear space
$gl(N,\mathbb{C})$ of $N\times N$ complex matrices, and in
particular one can expand a matrix $U$ as

\begin{equation}
U=\frac{1}{N^2}\sum_{k\in\mathbb{Z}_N^2}U(k)J_k, \qquad
U(k)=N\mathrm{tr}_{(N)}\left(UJ_k^\dag\right).
\end{equation}
We can interpret the momentum coefficients as the dynamical
degrees of freedom in the TEK model.\\
In analogy with the continuum counterpart (\ref{delta}) we can
define the operator (this time on a discrete torus)

\begin{equation}
\Delta(x)=\frac{1}{N^2}\sum_{k\in\mathbb{Z}_N^2}J_k \ e^{-2\pi i
k_ix^i/L},
\end{equation}
where $L=a N$ is the dimensionful extension of the lattice with
$N^2$ sites $x^i$. Because of the relations (\ref{periodicity})
the $\Delta(x)$ matrices are Hermitian and periodic in $x^i$ with
period $L$, and thus the lattice is a discrete torus.

In analogy with the continuum formalism depicted in section
\ref{sezformalismo}, we can define an invertible map between
$N\times N$ matrices and lattice fields. Namely, we have the
following properties

\begin{equation}
\begin{array}{r c l}
\mathrm{tr}_{(N)}\left(J_k\Delta(x)\right) & = &
\frac{1}{N}e^{2\pi i k_ix^i/L},\\
\\
\frac{1}{N}\sum_x\Delta(x)_{\mu\nu}\Delta(x)_{\lambda\rho} & = &
\delta_{\mu\rho}\delta_{\nu\lambda},\\
\\
\frac{1}{N}\mathrm{tr}_{(N)}\left(\Delta(x)\Delta(y)\right) & = &
N^2\delta_{x,y(\mathrm{mod} \ L)}.
\end{array}
\end{equation}
which yield a natural definition for the lattice field
$\mathcal{U}(x)$ from the Fourier modes of its matrix partner $U$:

\begin{equation}
\mathcal{U}(x)\equiv\frac{1}{N}\sum_{k\in\mathbb{Z}_N^2}U(k)e^{2\pi
i k_ix^i/L }=\frac{1}{N}\mathrm{tr}_{(N)}\left(U\Delta(x)\right).
\end{equation}
Since
\begin{equation}
U=\frac{1}{N^2}\sum_x\mathcal{U}(x)\Delta(x),
\end{equation}
the unitarity condition on the matrix $U$ is translated on the
field $\mathcal{U}(x)$ in terms of $U(1)$ \emph{star unitarity}:

\begin{equation}
\mathcal{U}(x)\star\mathcal{U}^*(x)=\mathcal{U}^*(x)\star\mathcal{U}(x)=1,
\end{equation}
where the \emph{lattice star product} is defined by the natural
discrete analog of eq. (\ref{contstarprod}), namely

\begin{equation}\label{starprod}
\begin{array}{r c l}
\mathcal{A}\star\mathcal{B} & \equiv &
\frac{1}{N}\mathrm{tr}_{(N)}\left(AB\Delta(x)\right)\\
\\
 & = &
 \frac{1}{N^2}\sum_y\sum_z\mathcal{A}(x+y)\mathcal{B}(x+z)e^{2i(\theta^{-1})_{ij}y^iz^i},
\end{array}
\end{equation}
with dimensionful non-commutativity parameter

\begin{equation}\label{dimnoncom}
\theta_{\mu\nu}=\frac{a^2 N}{\pi}n_{\mu\nu}.
\end{equation}
The star product (\ref{starprod}) reproduces the continuum version
of eq. (\ref{contstarprod}) in the limit $a\rightarrow 0$, and it
reproduces the same algebraic properties with space-time integrals
replaced by lattice sums.

In order to reproduce the non-commutative $U(1)$ theory, we
substitute the completeness relation

\begin{equation}\label{completness}
\frac{1}{N^2}\sum_x\Delta(x) \ = \ \mathbb{I}_N
\end{equation}
into the action (\ref{azioneTEK}) and obtain

\begin{equation}
S_{TEK}=-\frac{\beta}{N}\sum_x\sum_\mu
\mathrm{tr}_{(N)}\Big[\left(U\Gamma_\mu U^\dag\Gamma_\mu^\dag \ +
\  h.c.\right)\Delta(x)\Big].
\end{equation}
As in the context of twisted reduced models, the matrices
$\Gamma_\mu$ act as lattice shift operator, and thus they behave
as discrete derivatives $e^{a\hat{\partial}_\mu}$. Indeed one can
show from above that

\begin{equation}
\Gamma_\mu\Delta(x)\Gamma_\mu^\dag=\Delta(x-a\hat{\mu}),
\end{equation}
from which it follows that shifts on the fields are represented as

\begin{equation}
\mathcal{U}(x+a\hat{\mu})=\frac{1}{N}\mathrm{tr}_{(N)}\left(\Gamma_\mu
U\Gamma_\mu^\dag\Delta(x)\right).
\end{equation}
Therefore, we can rewrite the action (\ref{azioneTEK}) as

\begin{equation}\label{azioncinafinale}
S_{TEK}=-\beta\sum_x\sum_\mu
\Big[\mathcal{U}(x)\star\mathcal{U}^*(x+a\hat{\mu})+\mathcal{U}(x+a\hat{\mu})\star\mathcal{U}^*(x)\Big].
\end{equation}

\subsection{The non-commutative theory}

The theory described in eq. (\ref{azioncinafinale}) is naturally
$U(1)$ left and right invariant, i.e., given a constant field
$g\in U(1)$ the action is invariant under the transformations
(where the ordinary product is intended)

\begin{equation}
\mathcal{U}(x)\rightarrow g \cdot\mathcal{U}(x),\qquad
\mathcal{U}(x)\rightarrow \mathcal{U}(x)\cdot g.
\end{equation}
Let us turn to the commutative continuum version of the theory,
whose field we call $u(x)$. First of all we perform the following
substitution, dictated by the $U(1)$ unitarity condition

\begin{equation}
u(x)u^*(x)=1  \ \rightarrow \ u(x)=e^{i\varphi(x)},\qquad
\varphi\in\mathbb{R}.
\end{equation}
The action then reads, up to the coupling constant

\begin{equation}
S=\int\mathrm{d}^2x \
\partial_\mu\Big[e^{i\varphi(x)}\Big]\partial^\mu\Big[e^{-i\varphi(x)}\Big]=\int\mathrm{d}^2x \
\partial_\mu\varphi(x)\partial^\mu\varphi(x),
\end{equation}
and therefore the theory is equivalent to the theory of a free
massless real field.

Turning now to the non-commutative continuum theory of eq.
(\ref{fieldaction}), we notice that if the field $\mathcal{U}(x)$
decreased sufficiently rapidly at infinity we could integrate by
parts and turn the star product into the standard one. The \emph{action}
would thus be the same as in the commutative version. Naturally,
the field is subject to the constraint of star $U(1)$ unitarity,
which reads

\begin{equation}\label{condisiun}
\begin{array}{c}
\mathcal{U}(x)\star\mathcal{U}^*(x)= \ 1 \ =\\
\\
\mathcal{U}(x)\mathcal{U}^*(x)+\sum_{n=1}^\infty\left(\frac{i}{2}\right)^n
\frac{1}{n!}\theta^{i_1j_1}\cdots\theta^{i_nj_n}
\partial_{i_1}\cdots\partial_{i_n}\mathcal{U}(x)\partial_{j_1}\cdots\partial_{j_n}\mathcal{U}^*(x).
\end{array}
\end{equation}
This condition naturally implies that the theory is not a free
theory as in the commutative case, although the action would be
formally the same. Moreover it is not clear if a scalar complex
field subject to the condition (\ref{condisiun}) can at the same
time satisfy the rapidly decreasing condition needed to integrate
the non-commutative action by parts and neglecting the boundary
behavior. Therefore, we will still use as the reference action of
the continuum theory we wish to study eq. (\ref{fieldaction}).

\subsection{Physical interpretation: the brane vacuum}

The low energy effective action for a $p$-brane in the presence of
nonzero constant $B_{\mu\nu}$ field along the brane is given by
the dimensional reduction of the 10-dimensional non-commutative
Yang-Mills model to the brane world volume \cite{Connes}. In
particular, in the limit of large field $B_{\mu\nu}$, i.e. when
\begin{equation}\label{limitt}
\alpha^\prime\|B_{\mu\nu}\|\gg\|g_{\mu\nu}\|,
\end{equation}
the non-commutativity parameter $\theta_{\mu\nu}$ is given by
\begin{equation}\label{constfield}
\theta^{\mu\nu}=(B^{-1})^{\mu\nu}.
\end{equation}
In the case of a brane-antibrane pair of a non-BPS non-stable
brane, one finds in the spectrum of the effective theory also
tachyonic modes, described by non-commutative scalar fields $T$
with tachyonic potential $V(T)$ \cite{tachiun}.\\
In the trivial non-commutative gauge field background, the part of
the action of the brane system containing the tachyonic mode is
given by the same action describing a non-commutative Higgs-like
model of charged scalar fields, i.e.:
\begin{equation}
S=\int
\mathrm{d}^{p+1}x\left(\frac{1}{2}\partial_\mu\phi\star\partial^\mu\phi^\dag
- V(\phi\star\phi^\dag)\right),
\end{equation}
where $V(\cdot)$ is a potential with a nontrivial v.e.v.:
$|\phi|^2$= some constant, and the field $\phi$ transform in the
bi-fundamental representation of the $U(1)$ gauge group
\cite{harvey}.\\
A point in the (true) vacuum of the field $\phi$ can be
parametrized by an element of the non-commutative $U(1)$,
\begin{equation}\label{transofrmations}
\begin{array}{r c l}
\phi & \rightarrow & U\star \phi\\
\phi^\dag & \rightarrow & \phi^\dag\star U^{-1},
\end{array}
\end{equation}
where $U$ and $U^{-1}$ satisfy $U\star U^{-1}= U^{-1}\star U =
1$.

To get the action describing the dynamics along the valley of the
potential $V$ in terms of the field $U$, one has to take $\phi$
constant at the minimum of the potential, perform the
transformation (\ref{transofrmations}) and declare $U$ dynamical
\cite{Sochi}. The action for the Goldstone $U$-\emph{field} is then
\begin{equation}
S=\frac{1}{\lambda^2}\int
\mathrm{d}^{p+1}\eta^{\mu\nu}\partial_\mu U\star\partial_\nu
U^{-1},
\end{equation}
where $1/\lambda^2=\phi\star\phi^\dag$.

Our model then corresponds to the case $p=1$, with Euclidean
metrics, and with the obvious identifications for the couplings
$\lambda,\beta$ and for the constant field $B_{\mu\nu}$ and the
non-commutativity parameter of eq. (\ref{dimnoncom}) according to
(\ref{constfield}).

\subsection{Double scaling limit}

From (\ref{dimnoncom}) we see that in order to take the continuum
limit of the model in such a way that the dimensionful
non-commutativity parameter $\theta$ (which in the present case is
just a real number) is fixed, we must fix the quantity $a^2N$. It
is clear from above that we have to send $N$ to infinity if we
want $a$ to go to zero and the dimension of the lattice to go to
infinity. In order to set $a\rightarrow 0$ we have to tune somehow
the coupling constant $\beta$. From renormalization group analysis
of the beta function of chiral models it is known that

\begin{equation}\label{relazz}
a\sim \Lambda^{-1}e^{-c \beta}, \qquad c=8\pi.
\end{equation}
It is questionable whether a relation like (\ref{relazz})
is valid in the context of the TEK reduced model, because whether the equivalence
between principal chiral models and TEK reduced models holds is in itself
a nontrivial question \cite{future}. Moreover
equation (\ref{relazz}) is strictly valid only in the planar
limit. Nevertheless, we would like to propose to \emph{assume}
such a relation, and to numerically verify its consistency.

An analogous assumption is made in \cite{nishi}, where the relation
between $a$ and $\beta$ is taken to be the known Gross-Witten
planar result. The numerical results presented in \cite{nishi}, incidentally, 
strongly confirm the validity of such hypothesis.

Whether a different value for $c$ from the one
indicated in (\ref{relazz}) or a different functional dependence would lead to
similar results to what we present in sec. \ref{sesiun} is a non trivial legitimate question.

What we propose to do is therefore to send $N$ and $\beta$ to
infinity in such a way that $\vartheta\equiv N\cdot
e^{-16\pi\beta}$ is kept fixed. Numerical analysis indicates that
indeed finite $N$ and $\beta$ effects tend to compensate in such a
limit, in a manner similar to the one obtained in \cite{nishi} for
the $2D$ EK model.

\subsection{Correlation Functions}

Typical objects that can be studied in numerical simulations are
correlations functions. In particular, the easiest one to compute
turns naturally out to be the two points function. In particular,
given a lattice site $n=(n_1,n_2)$, we define, in the reduced
model, the following function, which is nothing but the translated
version of

\begin{equation}
G(\mathbf{n}) \ \equiv \ \frac{1}{N}\langle\mathbb{R}\mathrm{e} \
\mathrm{tr}_{(N)} \ \left(U(\mathbf{n})U(0)^\dag\right) \rangle
\end{equation}
via the substitution of eq. (\ref{gammini}), namely

\begin{equation}
G_{TEK}(\mathbf{n}) \ \equiv \
\frac{1}{N}\langle\mathbb{R}\mathrm{e} \ \mathrm{tr}_{(N)} \
\left(\Gamma_1^{n_1}\Gamma_2^{n_2}U(\Gamma_2^\dag)^{n_2}(\Gamma_1^\dag)^{n_1}U^\dag\right)
\rangle.
\end{equation}
Again substituting the completeness relation (\ref{completness}) we
get

\begin{equation}
G_{TEK}(\mathbf{n}) \ = \ \frac{1}{N^2}\sum_x
\langle\mathbb{R}\mathrm{e}
\left(\mathcal{U}(x+\mathbf{n})\star\mathcal{U}^*(x)\right)\rangle.
\end{equation}
Since the non-commutative theory is defined on a lattice with $N^2$
sites, this expression defines the average of the two point
function over all the possible lattice sites, and thus gives a
coherent expression for the two point function of the theory.

Incidentally, the internal energy of the model is given, up to
constants, by \mbox{$G_{TEK}(1,0)+G_{TEK}(0,1)$}.

\section{Numerical Results}\label{sesiun}
What should we expect from numerical Monte Carlo analysis? First
of all, if the double scaling limit we take is correct, we should
expect a coherent superposition of the behavior of the
correlation functions as $N,\beta\rightarrow\infty$. If this is
the case, it would strongly indicate the correctness of the
procedure. Secondly what we should not expect is the typical
behavior of the finite $N$ corrections found in the reduced model
in the strong coupling regime \cite{future}. In the double scaling
limit one takes into account also the non-planar diagrams, and it
is therefore not clear how the finite $N$ and $\beta$ effects
could
manifest.


\TABLE[ht]{
\caption{Set of values of $(N,\beta)$ at fixed $\vartheta$ which
we investigated numerically }
\label{tabella}
\begin{tabular}{| c ||  c | c |}

\hline

 & $N$ & $\beta$ \\
\hline\hline
$\vartheta_1=9.93\cdot10^{-6}$ & 60 & 0.31063\\
 & 70 & 0.31370\\
 & 80 & 0.31635\\
 & 100 & 0.32079\\
 & 120 & 0.32442\\
\hline\hline
$\vartheta_2=1.14\cdot10^{-6}$ & 50 & 0.35000\\
 & 80 & 0.35935\\
 & 100 & 0.36379\\
 & 120 & 0.36742\\

\hline

\end{tabular}
}

We used a Metropolis algorithm to update the $N\times N$
matrix $U$. Trial matrices were selected by multiplying
the actual matrix $U$ by a random $SU(2)$ matrix embedded in $SU(N)$,
choosing randomly among the $N(N-1)/2$ $SU(2)$ subgroups, and a $U(1)$ random phase. 
More precisely, once the $SU(2)$ subgroup and the $U(1)$ phase were randomly chosen, 
we performed ten Metropolis hits with an approximate
acceptance of 50\%.
Each $SU(2)$ updating requires $O(N)$ operations. 
The number of $SU(2)$-subgroup updatings per run was $O(10^7)$.
We should also mention that for the values of $N$ we investigated (see Tab. \ref{tabella}),
we observed some problem of thermalization
when using completely random 
configurations (for example, constructed  by
multiplying $N(N-1)/2$ completely random SU(2) matrix  embedded in
a $N\times N$  and associated to different subgroups, times a phase)
as starting point of our simulations.
In particular, we noted a worsening with increasing $\beta$.  So,
as starting point we used either moderately random matrices or
the unity matrix.
It is well known that a simple Metropolis algorithm
does not provide a particularly efficient method
to simulate a statistical system. Our choice was essentially due to
the fact that the reduced action is quadratic in the
matrix variable $U$. Moreover, it does not lend itself
to a linearization by introducing new matrix variables,
as in the case of the reduced TEK gauge theory \cite{FH}.

We numerically studied the theory for two particular values of
$\vartheta$. Our choice was motivated by two facts. First of all it
seems unnecessary to test the existence of a double scaling limit
in the strong coupling region (i.e. for
$\beta\lesssim\beta_c\approx0.3058$) since the reduced theory is
actually under control in that regime, and it has been shown
\cite{future} to accurately reproduce the standard, commutative parent
theory. Secondly, we must address to sufficiently high values of
$N$, because on the one hand the limiting procedure is expected to
be sensible in the large $N$ limit, and on the other hand the dual
lattice of the reduced theory has dimensions proportional to $N$.
Therefore, in order to avoid what we can legitimately call \emph{finite
size effects}, we had to resort to high values of $N$ (the correlation
length of the non reduced model is of the order of some lattice spacing in the region we investigated).\\
We choose to take the values reported in Tab. \ref{tabella}, and
to take $N\gtrsim50$. A reasonable statistics for our Metropolis
algorithm limited the highest $N$ value to $N\lesssim120$.

The correlation functions we studied was defined on the lattice as
\begin{equation}
\mathcal{G}(n) \ \equiv \
\frac{G_{TEK}(n,0)+G_{TEK}(0,n)}{2}.
\end{equation}
In Fig. \ref{fig1} and \ref{fig2}
we show our numerical results.

For large $n$, we see that asymptotically the obtained
values for $\mathcal{G}(n)$ agree
within error-bars.  We should mention that we 
numerically studied also the so called \emph{diagonal} two points correlation 
function on the lattice, defined by
\begin{equation}
\mathcal{G}_d(\sqrt{2}\cdot n) \ \equiv \ G_{TEK}(n,n).
\end{equation}
The results we obtained for $\mathcal{G}_d(n)$ show an analogous asymptotic superposition 
within error-bars for large $n$ in the double scaling limit, with values coherent with 
what was found for $\mathcal{G}(n)$.  In fact, the lattice definition of the two point correlation function
should not depend on either of the two definition one takes, and this can be viewed as 
another confirmation of the validity of the outlined scheme.

In all cases, what we find supports the expected
numerical scenario, and confirms that the procedure described above
indeed yields a sensible nonperturbative definition of the
non-commutative theory we described.

\FIGURE[ht]{
\epsfig{file=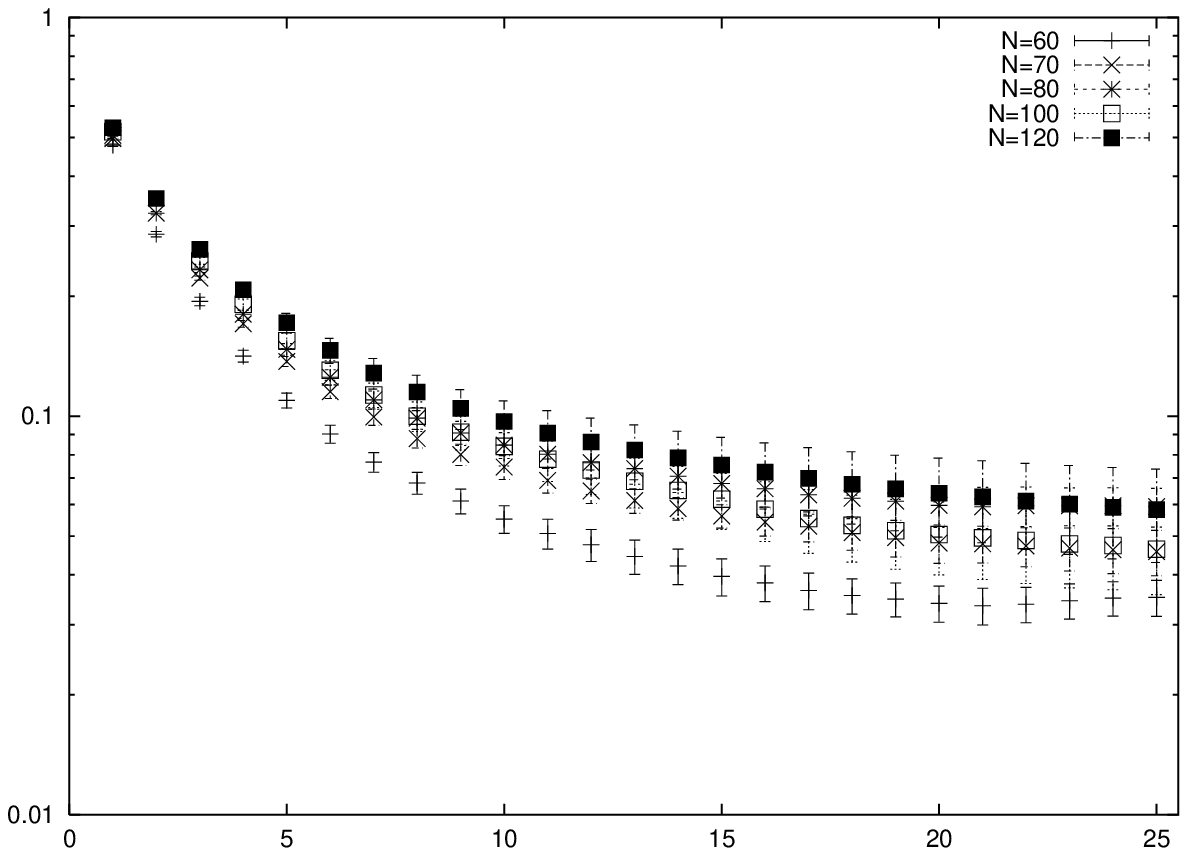, height=6.5truecm}
\caption{$\vartheta_1$: Numerical results for
$\mathcal{G}(n)$}
\label{fig1}
}
\FIGURE[hb]{
\epsfig{file=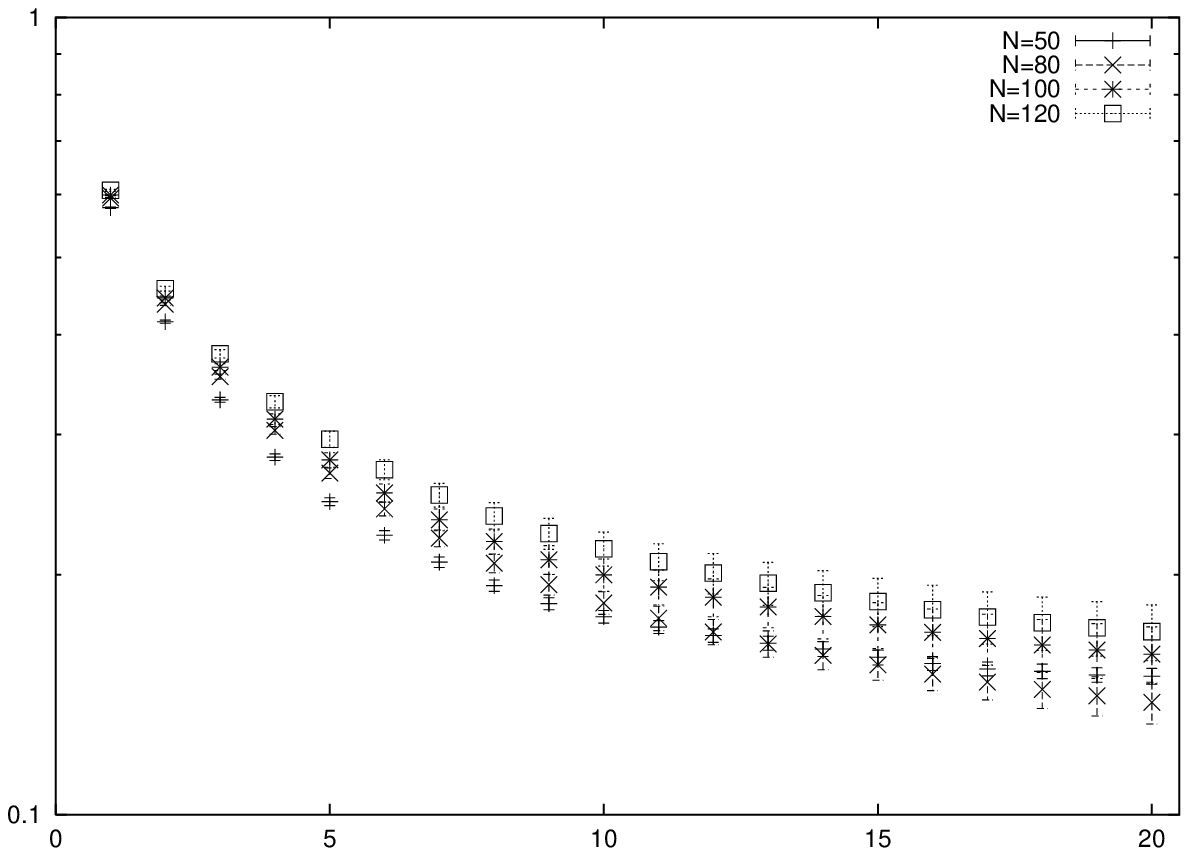, height=6.5truecm}
\caption{$\vartheta_2$: Numerical results for
$\mathcal{G}(n)$}
\label{fig2}
}

\clearpage

\acknowledgments{I warmly thank E. Vicari for helpful and useful discussions,
suggestions and comments. I also thank C. Sochichiu and M. Tsulaia for having drawn my
attention to their paper \cite{Sochi}.
} 


\end{document}